\documentclass[12pt,preprint]{aastex}

\slugcomment{Submitted to ApJL.}

\shorttitle{Polarimetry of blazar CGRaBS J2 0211+1051}
\shortauthors{chandra et al.}
\usepackage{graphicx}

\begin{document}


\title{Optical polarimetry of the  blazar CGRaBS~ J0211+1051 from MIRO}


\author{Sunil Chandra, Kiran S. Baliyan, Shashikiran Ganesh \& Umesh C. Joshi }
\affil{Astronomy \& Astrophysics Division, Physical Research Laboratory,
    Ahmedabad-380009, India}

\email{baliyan@prl.res.in}

\begin{abstract} We report the detection of high  polarization in the first detailed optical linear
polarization measurements  on the BL Lac object CGRaBS J0211+1051, which flared in $\gamma$-rays on
2011 January 23 as reported by Fermi. The observations were made during 2011  January 30 - February
3 using photo-polarimeter mounted at the 1.2m telescope of Mt Abu InfraRed Observatory(MIRO).  The
CGRaBS J0211+1051 was detected to have $\sim21.05\pm 0.41\%$ degree of polarization (DP) with steady
position  angle (PA) at 43$^\circ$ on 2011 January 30.  During Jan 31 and Feb 1, while polarization
shows some variation, position angle remains steady for the night.  Several polarization flashes
occurred during February 2 and 3 resulting  in changes in the DP by more than 4\% at short time
scales ($\sim$ 17 to 45 mins).  A mild increase in the linear polarization with frequency is noticed
during the nights of February 2 \& 3.   The source exhibited  significant  inter-night variations in
the  degree of polarization (changed by about 2 to 9\%) and position angle (changed by 2 to
22$^\circ$)  during the five nights of observations. The intra-night activity shown by the source
appears to be related to turbulence in the relativistic jet. Sudden change in the PA accompanied by
a rise in the  DP could be indicative of the fresh injection of electrons in the jet.  The detection
of high and variable degree of polarization categorizes the source  as low energy peaked blazar.

\end{abstract}

\keywords{Galaxies: active--- galaxies: nuclei--- techniques: polarimetric--- methods: observational--- BL Lacertae objects: individual (CGRaBS J0211+1051)}

\section{Introduction}

CGRaBS J0211+1051, also known as MG1 J021114+1051, 1FGL J0211.2+1049, 87GB 020832.6+103726 etc,  (RA 02:11:13.2, DEC 10:51:35, J2000)
 was found to have  a featureless optical spectrum and R-band magnitude of 15.42 in optical characterization of bright blazars in the uniform all sky  survey, CGRaBS \citep{Healey08}.  It was identified as radio source  and/or BL Lac object in 8.4GHz survey of  bright, flat-spectrum radio sources \citep{Healey07} and  in several other surveys,  including  4775 MHz  survey \citep{Lawrence86}. \citet{Snellen02} did optical identification of the radio sources from Jodrell Bank-VLA astrometry survey with $S_{6cm} > 200 mJy$ and reported R-band magnitude of 15.41 along with  384 mJy and 318 mJy flux values  at 5 GHz and 1.4 GHz, respectively. The source was detected in  $\gamma$-rays (E$> 100Mev$) by the Large Area Telescope (LAT) on board Fermi and is listed in the first Fermi catalog \citep{Abdo10}. Recently, the redshift of CGRaBS J0211+1051 was  estimated   to be z = $0.20\pm0.05$ from  its host galaxy observations \citep{Meisner10}.

On 2011 January  23,  flaring activity at $\gamma$-Rays (E$>100MeV$) was detected in  CGRaBS J0211+1051 by the  LAT on-board  Fermi  \citep{Ammando20}. The measured flux  at the level of 
$1.0 (\pm 0.3) \times 10^{-6} ph~cm^{-2} s^{-1}$ was 25 times the flux averaged over last 11 months \citep{Abdo10}. Swift/UVOT TOI   observations \citep{Ammando29}  on 2011 January 25 found CGRaBS J0211+1051 about 1.3 and 1.4  magnitude brighter in $U$-band ($13.77\pm0.03$ mag) and  W2-band ($14.44\pm0.04$ mag) respectively, as compared to their values on  2010 March 5. Swift also observed CGRaBS J0211+1051 in V band with $14.00\pm0.08$ mag.  On 2011 January  27, \cite{Nesci11}  reported $R_c$ band  magnitude of 13.37 for the source showing it to be 1.74mag  brighter than POSS~I red plate value  and  Digitized First Byurakan Survey (DFBS) value of  R=15.1 as obtained in 1950 and 1971, respectively. 
 
 \cite{Kudelina11}, based on their past observations made with MASTER-net \citep{Lipunov04}, found the   source to have continuously increased in brightness from 14.45 mag in white light on 2010 February 3 to 13.35   mag on 2011 January 24. The unfiltered magnitudes in their observations were estimated based on 3-comparison stars  in the USNO B1.0 catalogue. \cite{Dorgovski11} report a steady rise  from ~15.6 mag in 2008, to ~13.5 mag on 24 Jan 2011, with superposed  variations on a  month time scale with amplitudes as large as 0.5 - 1 mag. The source also showed two sharp flares, in mid-Nov. 2006 (peak magnitude 14.2) and in early September 2007 (peak magnitude 14.4). All these reports suggest that CGRaBS J0211+1051 was in brightening phase for last several  years with short (months) time scale variations. The amplitude of variation (0.5 - 1 mag) on a time scale  of months suggests strong activity in the source in optical while  it is flaring in high energy $\gamma$-rays.  It would be  interesting to see the behaviour of this  source on intra-night time scales with continuous monitoring.

 Following the flaring of the source in $\gamma$-rays, \cite{Gorbovskoy11} made optical polarimetric  measurements in V-band using MASTER-net observational sites at Tunka-Baykal and Amur-Blagoveschensk and found  the source to have 12 \% polarization on 2011 January 28. The corresponding V-band brightness magnitude of the source  was about 13.65 mag. \cite{Sunil11} detected even higher degree of polarization ($>$21\%) in white light during their observations from Mt Abu InfraRed Observatory (MIRO) on 2011 January 30. 
 
Extreme variations in the flux and polarization at various time scales across the whole electromagnetic spectrum are the characteristics of the blazars, a subclass of AGNs seen at small angle ($\leq 10^\circ$) to the jet emanating from very close to the black hole \citep{Urry95, Bland79}. Such variations could be caused by the perturbations in accretion disk or relativistic jet as descibed by several models \citep[e.g.][]{Mw93, Marsg85, Qi91, Mar96,  Gopalw92}.

Since radio to x-ray emission in blazars can be associated with synchrotron radiation, systematic polarization measurements provide important tool to understand the nature of such variations and help constrain the models of emission. A detailed review of polarization properties of the blazars is given by \citet{Angel80}. The study of variation in polarization  is also useful in probing the structure of the jet and the nature of physical processes in AGNs \citep{Mars06, Andru05, Villforth09}. Motivated by this, we made detailed optical  polarization measurements  from the MIRO  during 2011 January 30 to February 3. The main objective  being to investigate the day-to-day variability in polarization and position angle and any possible intra-night  activity in  the source. These are, to our belief, first detailed and systematic polarization data on the   CGRaBS J0211+1051 reported so far.

\section{Observations \& Data Analysis}

The observations were made using the two channel PRL Photo-polarimeter (PRLPOL) mounted on the 
Cassegrain focus of the  1.2-m
telescope of MIRO, operated by the Physical Research Laboratory, Ahmedabad, India during five
consecutive nights, 2011 January 30 to February 3. The PRLPOL, 
described in detail by \citet{Desh85}, was recently fully refurbished and automated\citep{Ganesh09}. 
It is a rotating half-wave plate polarimeter with
a Wollaston prism that divides the incident light beam into two
components, each one directed to a different photomultiplier tube. The
instrument has a $UBVRI$-system filter wheel and a second wheel with
diaphragms of different sized apertures. The observations were carried out
mostly in the white light to maximize the signal. For white light, the effective wavelength is 
determined by the sensitivity of the detector, here PMT (EMI 9863B) which peaks at $\sim \lambda=480$nm for CGRaBS J0211+1051.
Some measurements were made with B, V and R -bands to investigate any wavelength dependence of polarization. 
We adopted sky-source-sky observation strategy for the observations where, alternately, sky
and source were kept at the center of the aperture.  The sky measurements were taken about 30" away from the source. The exposure time for both, the sky and the 
source was kept 40 seconds during all the five nights for unfiltered white light observations and 120 secs for observations in B, V and R bands on 2011 February 2-3. The appropriate size of the aperture is 
chosen keeping in mind the optimum value of S/N ratio and the prescription of \cite{Andru08} for avoiding spurious variations caused by the possible change in the seeing and the contamination by the   host galaxy thermal  emission  which tends to decrease the value of intrinsic polarization. Too small ($\sim$ FWHM) an aperture will introduce spurious variations if the seeing changes while a large one would result in suppression of any  intrinsic variation and extent of the polarization of the source. In the present case, the host galaxy is more than three magnitude fainter than the source \citep{Meisner10} as the source is in relatively brighter phase \citep{Nesci11, Kudelina11}.  Based on these criteria we use  10 arcsec aperture  for the target and other stars used for the  calibration.  
Weather conditions were photometric with a moonless sky, which was more than two magnitude fainter than the source.

The degree of polarization and position angle are obtained from the online data reduction performed after each integration invoking a   least square fit to the counts from the two PMTs as a function of the position of the rotating half wave plate. The mean error in the polarization is estimated from the actual deviation of the counts from the fitted curve.  Standard stars were observed every night to determine the zero point for the position angle and the instrumental polarization, which was 
found to be negligibly small ($<0.02$).

\section{Results \& Discussion}

 The position angle of polarization (PA) for the source was corrected using measurements on the 9-Gem and  error in PA was calculated using the expression by \cite{Serkowski74}.  The nightly averaged values of DP and  PA were calculated and their standard deviations obtained. 
 In Table 1 we report polarization data during the  observing run giving date, MJD,  duration of observation in hours, nightly averaged values of the DP, PA  and their respective standard deviations.
The polarization data reported here are obtained through 10 arcsec aperture as mentioned in the earlier section.  The sky was very stable and more than two magnitude fainter than the source throughout the observations. A larger aperture might result in significant reduction in the DP due to the thermal emission from the host galaxy contaminating the non-thermal emission from the nucleus, particularly when the host galaxy is bright. In the present case, CGRaBS J0211+1051 and its host galaxy have I' band magnitudes of 15.32 and 16.91, respectively, as reported by \cite{Meisner10} during their 2008 October 31 to November 2 observations.  Since galaxy light peaks in the near IR band and has reduced brightness at shorter wavelength, the effect of galaxy light contaminating the nuclear emission will reduce in the optical B, V and R bands. Also, the source was in fairly bright phase during our observations \citep{Nesci11, Kudelina11} as compared to the 2008 level,  therefore we do not expect  any significant contamination of the nuclear light. Nevertheless, the polarization values reported here should be marginally lower than their intrinsic values.

The blazars are known to show large amplitude rapid variations in flux and polarization at various time scales. The nature of such variations can be used to infer the physical processes at work in these sources. To investigate the intra-night polarization behaviour of CGRaBS J0211+1051, DP and PA are plotted as a function time (MJD) in Fig.1(a-c) and Fig.2(a,b) for the observations in Jan 30 to Feb 1 (in white light) and Feb 2 \& 3 (in B,V, R bands \& white light), respectively. In the following we present polarization results from these observations on nightly basis.
 
On 2011 January 30, source was highly polarized at more than 21\% level during 20 mins of observations (Fig.1a).  The PA was  $\sim$ 43 deg. We do not notice any significant variation in DP or PA  during this night. However, this value of DP is much higher than the value reported by \cite{Gorbovskoy11}  on 2011 January 28 (DP=12\%), just two days prior to our measurements. We have 14 data points on January 31 obtained within 0.7 hrs of observation. At the onset of observations, DP is about 13\%, increasing to 13.8\% before falling by 2.7\% to 11.1\% within about 20 mins (decay rate $\sim$ 8\%/Hr). The DP starts increasing again and reaches about 13.2\% level in 18 mins time (rising rate 7\%/Hr) .  PA, however,  remains well behaved without any appreciable variation.
February 1 has large number of observational points (87) during 2.4 hrs of observation. Several microvariability events appear to be superimposed over a non-varying component (Fig. 1c). However, except for the  two events beginning at MJD 55593.424 and MJD 55593.446 with more than $2\sigma$ amplitude of variation in DP, all other events show $\sim 1\sigma$ variation. The position angle stays within 31$\pm$3 deg without any structures.
 
The situation is entirely different on February 2 and 3 when CGRaBS J0211+1051 shows considerable variation as shown in Fig.2(a,b).  During these two nights, in addition to white light, we also made observations in B, V and R filter bands to see any wavelength dependence in the DP and  PA.    On February 2, DP rises from 11.5\% (at time 4.392) to 13.5\% (at time 4.416) in white light (cf Fig.2a). The rise continues in B band observations reaching more than 15\% at MJD=55594.43. Beyond that,  DP decreases in V and R bands, partly perhaps due to the increase in wavelength and partly due to intrinsic variation in the polarization of the source. Towards the end, observations are again in white light and DP shows slight increase.  PA largely remains  within $\pm$ 3 degree range.
Fig.2b shows polarization  behaviour of CGRaBS J0211+1051 during February 3. Interestingly, the curve shows three quasi-periodic polarization flashes with significant amplitude of variation (up to ~ 4\%) through observations made with B, V, R bands and unfiltered white light. These events have fast rise and fall time scales ranging from 17 to 32 mins. PA varies between 41 \& 46 deg during the course of observations and can be considered as mildly variable. 

February 2 and 3 measurements with B, V and R bands are  indicative of the trend that degree of polarization increases with frequency in BL Lacs. The DP in B-band shows clear increase over R and V band  values (Fig 2) which can not be only due to the intrinsic variation as variation time scales are expected to be longer than  the temporal resolution (2 mins) used here.

 Now let us look at the Inter-night variations during 2011 Jan 30 to Feb 3. The Table 1 and Fig.3 show avaraged DP(\%) and PA($^\circ$) for all these nights. The error bars in the figure reflect the spread ($1\sigma$) due to intra-night variations in addition to the measurement errors. It is evident from the observed data that source was highly polarized on January 30 with DP about 21\% which decreases by 9\% and 11\% on Jan 31, and Feb 1, respectively. It then starts increasing again reaching 15.5 \% on Feb 3. The PA also changes significantly from night to night, initially following the changes in DP but dropping to 45$^\circ$ on the last night while DP increases to more than 15\%. We notice changes in PA by 2 to 22 deg while remaining within 28 to 53$^\circ$ range during our observations.

Apart from visual inspection of the polarization curves to look for variations, we also carried out statistical analysis to detect and quantify the variation parameters using the criterion of \citet{Kesteven76} applied by several authors to the variability
studies \citep[eg][]{Romero94}. Here, the variability in DP and PA is described by  the fluctuation
index $\mu$  and the fractional variability index of the source $FV$ obtained from the individual night's data.  The corresponding  expressions are:
\begin{equation} \mu = 100 \frac{\sigma_{\rm S}}{\langle
S \rangle}\, \;  {\%}, \end{equation}

\begin{equation}
FV = \frac{S_{\rm max} - S_{\rm min}}{S_{\rm max} + S_{\rm min}},
\end{equation}

where $\sigma_S$ is the standard deviation, $\langle S
\rangle$ is the mean value of the DP or the PA
obtained during the particular night, $S_{\rm max}$ and $S_{\rm min}$ are,
respectively, the maximum and minimum values for the DP or  PA. The source is classified as variable if the probability of exceeding the
observed value of

\begin{equation}
x^{2} = \sum_{i=1}^n{\epsilon_{i}^{-2}\, (S_i-\langle S
\rangle)^{2}}
\end{equation}
by chance is $<$ 0.1 \%, and non-variable if the probability is
$>$ 0.5 \%. For in between values of p($x^2$), source can be taken as possibly variable (PV). In the above expression, $\epsilon_i$ are the  uncertainties in the individual measurements.
If the errors are random, $x^2$ should be distributed as $\chi^2$ with $n-1$ degrees of freedom, where $n$ is the number of points in the distribution.

Table 2 shows the values of the variability parameters
for the DP and PA.  Columns 1 \& 2 give the  Date and the number
of observation points, Columns 3 -5  present the values of $\mu$,  $FV$  and $\chi^2$ for DP, and Column 6 gives the status of the source (V: Variable; NV: Non-variable; PV: Possibly Variable). Rest of the 4-columns give values of   $\mu$, FI,  $\chi^2$ and variability status for PA.  These results quantitatively substantiate significant intra-night variability during February 2 \& 3 in the polarization behaviour of the source. 

Let us briefly  discuss these results.\\
 The observed optical emission in blazars originates in a part
of the accretion disk and the inner (pc-scale) regions of the jet. Thus, one can discuss the possible reasons
behind the variations in the degree of polarization and position angle over various time scales.  The polarization caused by the electron or dust scattering in the accretion disk is usually low (few percent). Since CGRaBS J0211+1051 shows high DP (10-21\%) during present observations, emission must be dominated by the relativistic jet, aligned at a small angle to the line of sight. This emission is synchrotron radiation from the relativistically moving electrons  in the jet and is highly polarized ($>$ 70\%) if the magnetic field is uniformly aligned. Reduced observed DP indicates a chaotic magnetic field, which can be described in terms of N cells  with uniform field but randomly oriented. The degree of polarization is also reduced by geometrical depolarization due to variation of the magnetic field orientation along line of sight and the contamination by the thermal emission from host galaxy.  The position angle is orthogonal to the projected direction of the magnetic field. However, relativistic motion aberrates the angle resulting in PA to be more aligned with the jet direction \citep{Mars06}.   In BL Lacs, the parsec scale magnetic field in the jet is tangled and shocks moving down the jet compress the magnetic fields, aligning it perpendicular to the flow direction \citep{Marsg85}.  The interaction of relativistic shocks with features in the pc-scale jet results in rapid variations in the flux and polarization \citep{Marsgt92, Qi91}. The features are varied in nature depending upon the model and generally are sub-pc in size. Macroscopic Kelvin-Helmholtz instabilities are capable of producing such features in the inner beam. Quasi-periodic variations could be caused by the regularly spaced obstacles in the path of the jet. Such models can explain variations with time scales of weeks to  days.  Faster variations down to the sub-hour time scales can not be explained due to limited thickness of the shocks.  The rapid, intra-night flickerings in the degree of polarization as observed during  2011 February 1-3 could be the effect of turbulence in the post-shock region of the jet.
 
The position angle of polarization suffered drastic changes between the nights of January 30 \& 31, February 2 \& 3 by $> 8^\circ$ and  $10^\circ$, respectively. The DP changed by more than and 8\% and 3\% during these periods. The DP changed by more than 9\% during January 28 (12\% as reported by \cite{Gorbovskoy11}) and Juanury 30 (21\%).  These sudden changes in the PA and DP might be due to fresh injection of plasma blobs in the jet on January 28 and February 2.  The shocks thus formed compress and enhance the magnetic field parallel to the shock front giving rise to sudden changes in the PA and DP. 

The extent and nature of linear polarization exhibited by CGRaBS J0211+1051 during 2011 Januray 30 to February 3 led us to infer that CGRaBS J0211+1051 is low energy peaked BL Lac which was in bright and active phase. Multifrequency observations are required to study its spectral energy distribution and determine the position of the two peaks.

\section{Conclusions}
First detailed optical polarization measurements are reported for the blazar CGRaBS J0211+1051 performed during 2011 January 30 to February 3.  The source shows high and variable degree of polarization ranging from ~21\% to 10\% during this period. We do not see significant intra-night variation during first three nights. However, substantial intra-night variability is seen on February 2 \& 3 with DP changing by more than 4\%. The position angle of polarization remains within 2$\sigma$ during the individual nights but changes significantly from night to night. The sudden changes in position angle could be indicative of the fresh injection of the shocks in the jet. The rapid intra-night flickerings in the  polarization appear to be due to small scale turbulence in the post-shock region of the jet.

There are no other polarization results in the literature for this source except for a report of 12\% polarization on 2011 Jan 28 by \cite{Gorbovskoy11}. The present results show a variation in DP from about 21\% to 10\% during 2011 January 30 - February 3 and therefore their value is in agreement with ours. The high value of polarization confirms the source to be a low energy peaked  BL Lac (LBL) object.  More multi-wavelengths observations, along with VLBI imaging, are needed to study the structure and  spectral energy distribution of the source to constrain the models of variability.

 This work is supported by the Department of Space, Government of India.

{\it Facilities:} \facility{MIRO:1.2m (PRLPOL)}


\clearpage


\begin{figure*}
\includegraphics[height=\textheight]{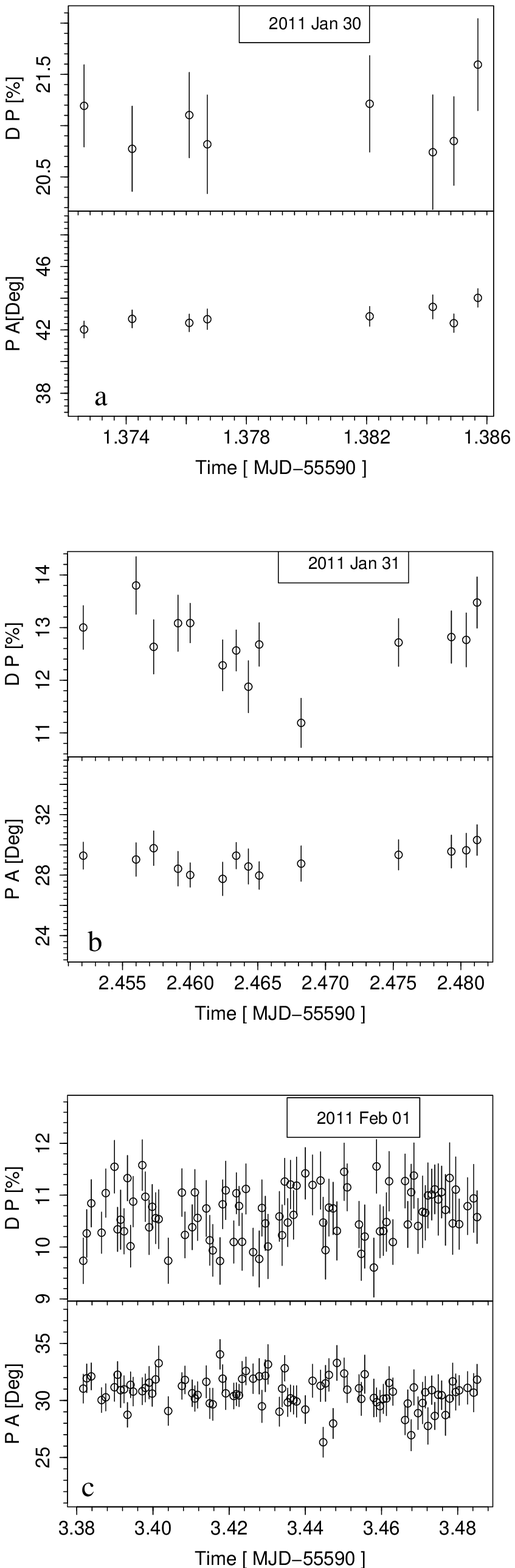}
\caption{Intra-night variations in the degree of  polarization (DP) and position angle (PA) during 2011
 January 30 to  February 1 for CGRaBS J0 211+1051 in white light.  The error bars show uncertainties in the individual measurements.}
\end{figure*}


\begin{figure*}
\includegraphics[width=\textwidth]{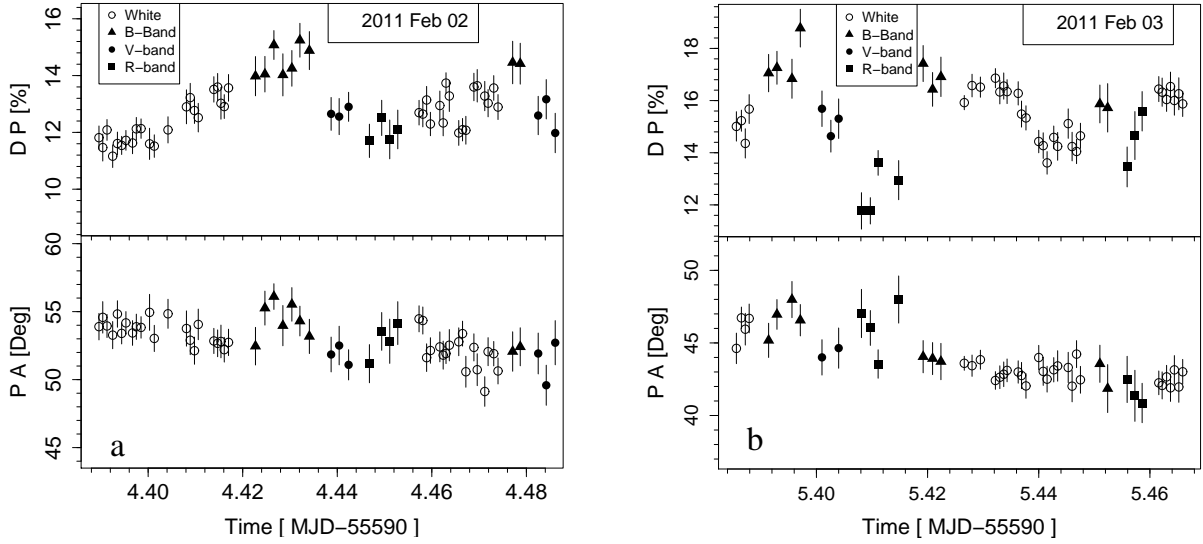}
\caption{Same as Fig.1 but for 2011 February 2 and 3 in B, V, R \& white light.}
\end{figure*}

\begin{figure*} 
\includegraphics{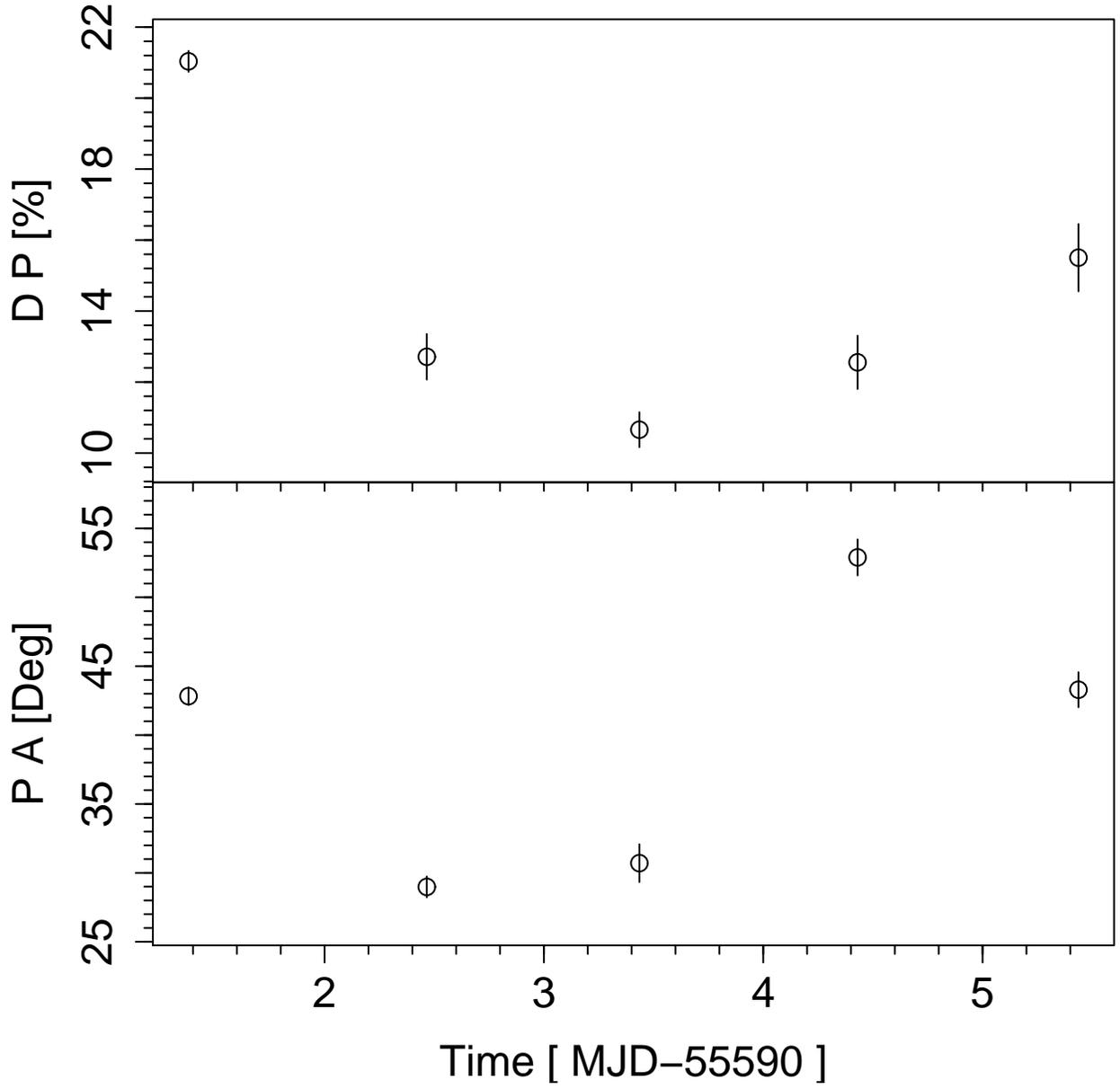}
\caption{Night to night variation in the DP and PA during 2011 January 30 to February 3.
 The error bars reflect the spread ($\pm \sigma$) due to intra-night variations.} 
\end{figure*} 
\clearpage 


\begin{table}
\caption{Nightly averaged polarization data for CGRaBS J0211+1051. Entries are: date of observation, MJD, duration of 
observation, degree of polarization (DP), $\sigma_{DP}$,  position angle (PA) and $\sigma_{PA}$. \label{tbl-1}}
\begin{tabular}{r c c c c  c   c}
\tableline \tableline
Date & MJD & $\Delta_T$(Hrs) & DP (\%)  & $ \sigma_{DP} $ (\%) &  PA($^{\circ}$) & $\sigma_{PA} ( ^{\circ})$  \\
\tableline
 2011 Jan 30      & 55591.3796	 & 0.336  & 21.052        & 0.295    & 42.771	& 0.634	\\
           Jan 31 & 55592.4665	 & 0.864  & 12.871	  & 0.489    & 28.963	& 0.758	\\
          Feb 1   & 55593.4352   & 2.498  & 10.643	  & 0.493    & 30.679	& 1.373	\\
         Feb 2    & 55594.4362	  & 2.112 & 12.629	  & 0.981    & 52.982	& 1.412	\\
         Feb 3    & 55595.4308	  & 1.920 & 15.481	  & 1.412    & 43.578	& 1.762	\\
\tableline
\end{tabular}
\end{table}

\begin{table}
\caption{Variability test results for CGRaBS J0211+1051. \label{tbl-2}}
\begin{tabular}{@{}rrrrrr@{~~~}r@{~~}r@{~}r r@{~~}r}
\hline
\noalign{\vskip 2pt}
\hline \noalign{\smallskip}
&& \multicolumn{4}{|c} {Degree of Polarization (DP)} & \multicolumn{4}{|c} {Position Angle (PA)} \\
~Date.& n &  \multicolumn{1}{|c}{$\mu(\%)$} & $FV$ & $\chi^2$ &  Status   &  \multicolumn{1}{|c} {$\mu (\%)$ }& $FV$ & $\chi^2$ &Status\\
\noalign{\smallskip}
\hline
\noalign{\smallskip}
2011 Jan 30 & 8   &  1.403 & 0.021  &  2.942 &  NV  & 1.482 & 0.023  & 8.215   & PV  \\
Jan 31      & 14  &  3.799 &  0.075 &  13.34 &   V   & 2.618 & 0.044 &  7.573  & NV  \\
 Feb 1      & 87  &  4.639 &  0.093 &  84.61 &  NV   & 4.475 & 0.127 & 89.691  & PV  \\
 Feb 2      & 58  &  7.774 &  0.154 &  213.20&   V   & 2.665 & 0.066 & 93.950  &  V  \\
 Feb 3      & 48  &  9.122 &  0.229 &  290.13&   V   & 4.042 & 0.080 & 120.411 &  V  \\

\noalign{\smallskip} \hline
\end{tabular}
\end{table}

\end{document}